# A new hybrid numerical scheme for modeling elastodynamics in unbounded media with near-source heterogeneities

Setare Hajarolasvadi and Ahmed E. Elbanna

Department of Civil and Environmental Engineering, University of Illinois at Urbana-Champaign, Urbana, IL 61801, USA. E-mail: hajarol2@illinois.edu ; elbanna2@illinois.edu

The Finite Difference (FD) and the Spectral Boundary Integral (SBI) methods have been used extensively to model spontaneously propagating shear cracks in a variety of engineering and geophysical applications. In this paper, we propose a new modeling approach, in which these two methods are combined through consistent exchange of boundary tractions and displacements. Benefiting from the flexibility of FD and the efficiency of spectral boundary integral methods, the proposed hybrid scheme will solve a wide range of problems in a computationally efficient way. We demonstrate the validity of the approach using two examples for dynamic rupture propagation: one in the presence of a low velocity layer and the other in which off-fault plasticity is permitted. We discuss possible potential uses of the hybrid scheme in earthquake cycle simulations as well as an exact absorbing boundary condition.

## 1. Introduction

Earthquake ruptures are nonlinear multiscale phenomena. The multiscale nature of the rupture process exists in both space and time. Spatially, a moderate size earthquake typically propagates over tens of kilometers. However, the physical processes governing the rupture propagation operates within a narrow region at the rupture tip, called the process zone, which may not exceed a few millimeters in size if realistic laboratory-based friction parameters are used (Noda et al. 2009). Temporally, an earthquake episode, where rapid slip occurs, only lasts for few to tens of seconds. However, the time required for stress buildup and the attainment of the right condition for the initiation of the friction instability during the interseismic period may be tens to hundreds of years (Lapusta et al. 2000). Thus, to resolve the full seismic cycle, it is necessary to design numerical protocols that capture spatial and temporal scales over several orders of magnitude. This is a fundamental challenge in earthquake source physics.

A breakthrough in addressing this challenge was achieved in the paper by Lapusta et al. (2000) using the spectral boundary integral technique. The boundary integral formulation enables reducing the spatial dimension of the problem by one, transforming 2D problems into 1D and 3D problems into 2D (Cochard & Madariaga 1994; Geubelle & Rice 1995). In that context, Lapusta et al. (2000) derived accurate adaptive time-stepping algorithms and truncation of convolution integrals that enabled, for the first time, the consistent elastodynamic simulation of a long sequence of events combining rapid slip during earthquake ruptures and slow deformation during the interseismic periods. Nonetheless, the method is only limited to homogeneous linear elastic bulk. While the method may be applied, in principle, to heterogeneous linear elastic materials, the lack of a closed form representation of the Green's function either inhibits the method from providing a well-defined solution to many problems of interest or makes it less computationally attractive. Furthermore, the superior performance of the spectral approach and its computational efficiency is only possible for planar interfaces. This precludes the representation of nonplanar faults or direct incorporation of fault zone complexity (e.g. damage, and shear bands)

On the other hand, numerical methods based on bulk discretization such as the finite difference and finite element methods have been used in simulating earthquake ruptures since mid-70s and early 80s with the pioneering works of Boore et al. (1971), Andrews (1976), Das and Aki (1977), Archuleta & Day (1980), Day (1982), Virieux & Madariaga (1982) and others. These methods are more flexible than the boundary integral approaches in handling heterogeneities, nonlinearities, and fault geometry complexities. Low-order formulations of these methods, however, suffer from some numerical problems, such as zero energy deformation modes and high frequency oscillations. Treatment of such artifacts requires adding artificial viscosity and hourglass control (Day et al. 2005). However, in recent years, highly accurate formulations were introduced, including the spectral finite element (Komatitsch & Tromp 1999), the discontinuous Galerkin method (Kaser & Dumbser 2006), and higher-order finite difference schemes (Kozdon et al. 2013). A main computational challenge of these methods is the need to discretize the whole bulk, which increases the computational demand by at least one order of magnitude compared to the boundary integral formulation. Furthermore, the computational domain must be truncated at a sufficient distance from the fault surface such that it would not affect the physical solution. This motivated the introduction of several widely-used absorbing boundary conditions such as boundary viscous damping (Lysmer & Kuhlemeyer 1969), perfectly matching layers (Berenger 1994), and infinite elements (Bettess 1977). However, in all these methods, artificial reflections exist to varying degrees and the absorbing surfaces must be taken sufficiently far from the fault surface to ensure solution accuracy. Moreover, attempts to perform cycle simulations using these volume-based methods are rare and have been restricted mainly to the quasi-dynamic limit (Erickson et al. 2016). This is partially due to the high spatial discretization cost and the lack of a systematic approach to handle both dynamic and quasidynamic calculations in the same framework which is required for simulating both earthquake ruptures and intersesismic slow deformations. Another challenge in these methods is defining fault loading. Currently, this is done by applying displacement-controlled loading at the far boundaries of the simulation box. This, however, makes the fault stressing rate dependent on where the domain is truncated. This problem is solved approximately in the spectral boundary integral formulation by loading the fault directly through back-slip.

Both bulk and boundary approaches have their merits and limitations. To that end, this paper proposes a novel hybrid numerical scheme that combines the finite difference method and the spectral boundary integral equation method to efficiently model fault zone nonlinearities and heterogeneities with high resolution while capturing large-scale elastodynamic interactions in the bulk. The main idea of the method is to enclose the heterogeneities in a virtual strip that is introduced for computational purposes only. This strip is discretized using a volume-based numerical method, chosen here to be the finite difference method for simplicity. The top and the bottom boundaries of the virtual strip are handled using the independent spectral boundary integral formulation (Geubelle & Rice 1995) with matching discretization. The coupling between the two methods is achieved as follows. The finite difference solution of the strip provides the traction to the spectral boundary method at each virtual interface. The spectral scheme is then used to predict the boundary displacements. These displacements are in turn applied to the strip to advance the solution to the next time step.

The remainder of this paper is organized as follows. In Section 2 we describe the model setup and the numerical methods (i.e. FD, SBI and the hybrid method). In Section 3, we investigate two problems to demonstrate the ability of the proposed numerical scheme in capturing the effects of nonlinearities and heterogeneities with no artificial wave reflections due to domain truncation. The first problem is a slip-weakening crack with a low-velocity fault zone and the second one is a slip-weakening crack with off-fault plasticity. Section 4 discusses the potential applications and some limitations of this new method as well as grounds for future work.

## 2. Model Setup and the Numerical Schemes

We consider a 2-D antiplane shear problem. The fault surface is chosen to coincide with the $x - z$ plane of a Cartesian coordinate system and all particles move in the $z$ direction while the rupture propagates along the $x$ direction. The only nonzero displacement component is $u_z(x, y, t)$, which satisfies the scalar wave equation in the case of a linear elastic homogeneous medium

$$\ddot{u}_z = c_s^2 \Delta u_z \tag{1}$$

Here, the dot denotes differentiation with respect to time; $c_s = \sqrt{\frac{G}{\rho}}$ is the shear wave speed and $G$ and $\rho$ represent the shear modulus and the density, respectively.

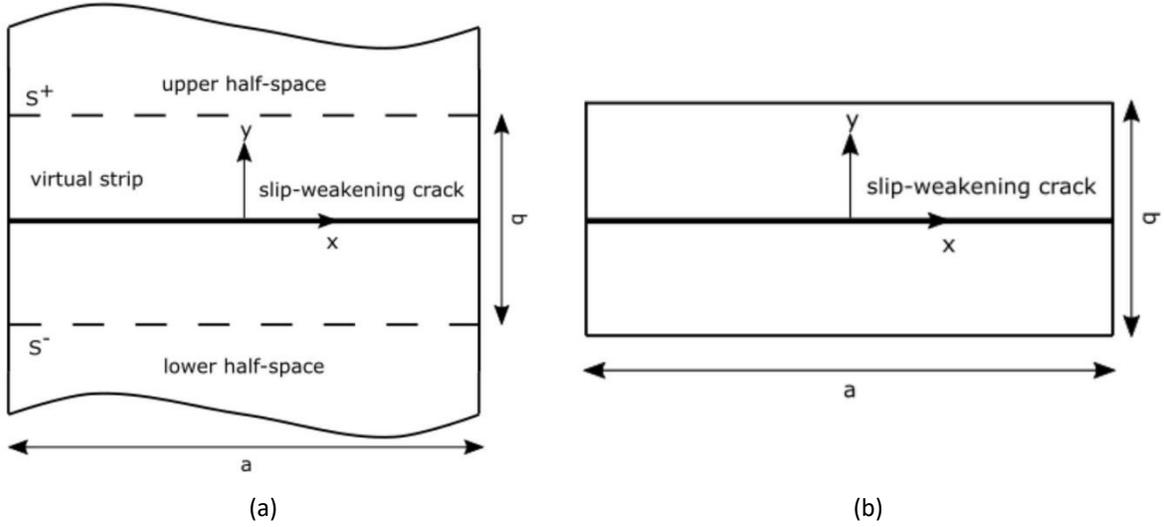

(a)                                (b)

Figure 1- Layout of the antiplane shear problem in the case of a planar fault coincident with the x-z plane. (a) General setup of the problem in the hybrid method. Two linear elastic homogeneous half-spaces bound a narrow domain, which exists in the vicinity of the fault and may be heterogeneous and/or nonlinear. (b) A sketch for the virtual strip that is introduced for computational purposes to isolate nonlinearities and/or heterogeneities for discretization using the finite difference method. The elastodynamic interactions through the bulk that influence the solution within the isolated domain are modeled by imposing the spectral boundary integral equations on the top S⁺ and bottom S⁻ edges of the virtual strip.

Slip is defined as $\delta u(x, t) = u_z(x, 0^+, t) - u_z(x, 0^-, t)$. The shear stress components of interest on the two half-planes of the fault are $\tau^+(x, t) = \sigma_{yz}(x, 0^+, t)$ and $\tau^-(x, t) = \sigma_{yz}(x, 0^-, t)$.

The fault constitutive response is governed by a linear slip-weakening friction law (Ida 1972; Palmer & Rice 1973) in which $\mu$ (the friction coefficient) is given by

$$\mu = \begin{cases} \mu_s - (\mu_s - \mu_d)\dfrac{\delta u}{D_c} & \delta u < D_c \\ \mu_d & otherwise \end{cases} \tag{2}$$

where $\mu_s$ and $\mu_d$ are coefficients of static and dynamic friction, respectively and $D_c$ is the critical slip-weakening distance. The friction coefficients are assumed to be spatially homogeneous along the fault. Table 1 specifies the values of the relevant parameters. The normal stress is equal to 100 MPa inside the nucleation zone and 120 MPa outside this region with the positive sign showing compression.

For nonlinear bulk, as in the case of simulations with off-fault plasticity, we will consider the following formulation for the elastodynamic problem

$$\begin{cases} v_z = \dot{u}_z \\ \rho \dot{v}_z = \tau_{xz,x} + \tau_{yz,y} \end{cases} \tag{3}$$

where $v_z$ is the particle velocity and $\tau_{xz}$ and $\tau_{yz}$ are the components of the shear stress. Given an appropriate nonlinear constitutive model relating the stress rate to the particle velocity gradient, a yield surface, and a flow rule, the above set of equations may be integrated. We provide specific details in Section 3.

Table 1- Stress and frictional parameters for the test problems. [Source: SBIEMLAB code script]

| Parameter | Symbol | Value |
|-----------|--------|-------|
| Initial shear stress , MPa | $\tau_0$ | 70 |
| Initial normal stress, MPa | $\sigma$ | 100 / 120 |
| Static friction coefficient | $\mu_s$ | 0.6 |
| Dynamic friction coefficient | $\mu_d$ | 0.33 |

## 2.1. Finite Difference Method

The finite difference scheme has been widely used since the 1970s to solve seismic wave propagation problems in elastic and inelastic media due to its simple formulation and easy implementation (Aochi et al. 2013 and references therein). In this paper, two formulations of the finite difference scheme are used. The first is a displacement formulation of the second-order wave equation on a non-staggered grid and the second is a discretization for the system of first-order hyperbolic equations (Eq. (3)) on a staggered grid using shear stress-velocity formulation following Day et al. (2005). The latter is used as it is more suited for implementing plasticity in the second problem since the constitutive relation evolves as part of the solution. The use of two FD formulations also attests the flexibility of the hybrid approach and its ability to allow for various numerical discretizations of the strip. To suppress zero energy modes and numerical oscillations in the second formulation, artificial viscosity and hourglass control are introduced in the model. A detailed discussion of the implementation of this method in the 2-D antiplane framework is given in Appendix A.

## 2.2. Independent Spectral Boundary Integral Method

In the spectral formulation of the boundary integral method, fault traction and slip are transformed into the Fourier domain. The formulation embodies an exact elastodynamic representation of the relation existing between the Fourier coefficients for the tractions and the corresponding displacement discontinuities (Geubelle & Rice 1995). There are two versions of spectral algorithms: in the first version, which is referred to as the independent spectral formulation, the elastodynamic response of each halfspace is formulated separately and then the two halfspaces are connected by imposing appropriate boundary conditions on the interface. In the second approach, referred to as the combined spectral formulation (Breitenfeld & Geubelle 1998; Lapusta et al. 2000), the formulation of the problem is written in a way to combine the elastodynamic analysis of the two halfspaces into one. The two approaches are based on the same principles and often yield similar results. However, they have minor differences in their formulation and implementation and show different stability characteristics when it comes to in-plane modes (Breitenfeld & Geubelle 1998).

We take advantage of the independent formulation in this paper since the boundaries of the virtual strip are not in the same geometrical location.

A thorough derivation of this approach for a 2-D antiplane fracture problem is given in Appendix B. The boundary shear stress and velocity are related by:

$$\tau_z^{\pm}(x,t) = \tau_z^{0\pm}(x,t) \mp \frac{\mu^{\pm}}{c_s^{\pm}} \dot{u}_z^{\pm}(x_1,t) + f_z^{\pm}(x,t) \tag{4}$$

The + and − represent upper and lower half planes. Here, $\tau_z^0(x,t)$ is the shear stress acting on the fault when it is locked and $f_z^{\pm}(x,t)$ represent the spatio-temporal convolution integrals of the elastodynamic Green's function and boundary velocities within the causality cones (See Eq. B24 and its counterpart for the lower half-plane for more details). This convolution term may be expressed in the Fourier domain as

$$f_z^{\pm}(x, y = 0^{\pm}; t) = F_z^{\pm}(t; q)e^{iqx} \tag{5}$$

where the Fourier coefficient $F_z^{\pm}(t; q)$ is obtained from Breitenfeld & Geubelle (1997) as

$$F_z^{\pm}(t;q) = \mp\mu^{\pm}|q| \int_0^t C_{III}(|k_n|c_s^{\pm}(t-t')) U_z^{\pm}(t';q)|q|c_s^{\pm}dt' \tag{6}$$

The convolution kernel of this independent formulation was shown to be $C_{III}(T) = \frac{J_1(T)}{T}$ with $J_1(T)$ as the first kind Bessel function of order one. This is identical to the convolution kernel of the combined formulation for the antiplane problem (Lapusta et al. 2000).

## 2.3. Hybrid Method

In the classical spectral boundary integral formulation, the shear stresses for the upper and lower half planes are coupled through the continuity of traction across the fault plane. Here, however, we consider cases where the linear elastic homogeneous bulk is not in the immediate vicinity of the fault surface. This may be the case, for example, if the fault is embedded in a low-velocity zone with elastic properties different from the host rock. Another example is rupture propagation on nonplanar fault surfaces or fault surfaces with coseismic damage and inelastic strain generation. In these cases, the direct application of the spectral boundary formulation is not possible due to either the existence of heterogeneity or invalidity of superposition because of nonlinearity. To take advantage of the homogeneity and linearity of most of the bulk, we propose a hybrid formulation as follows.

In the hybrid method, nonlinearities or heterogeneities are confined within a virtual narrow strip that also includes the fault or the wave source. This strip is discretized using a FD scheme in space and an appropriate integration scheme in time. The boundaries of the virtual strip are governed by the independent SBI formulation that represents the two elastic half-spaces outside the strip. Dirichlet and Neumann boundary conditions are imposed on the strip and the two half-spaces, respectively, at each time step to propagate the solution forward.

The general setup of the model for the hybrid method is shown in Figure 1. The width of the virtual strip depends on the nature of the problem but is always chosen such that the heterogeneities and any expected nonlinearities are fully contained within the strip. The boundaries of the virtual strip are taken in the linear elastic homogeneous bulk. A rectangular grid is introduced to discretize the strip with $N_x$ elements in the $x$ direction and $N_y$ elements in the $y$ direction. This grid is chosen to be non-staggered for the first test problem and partially staggered for the second one (see appendix A for the detailed configuration of the latter). The element size is defined as $h = \frac{\lambda}{N_x}$ where $\lambda$ denotes the domain size. The coordinates of the grid points are

$$x_i = i\,\Delta x \qquad i = -\frac{N_x}{2}, \dots, \frac{N_x}{2} \tag{7}$$

$$y_j = j\,\Delta y \qquad j = -\frac{N_y}{2}, \dots, \frac{N_y}{2} \tag{8}$$

Square-shaped elements are chosen in the strip to keep the error in the FD scheme uniform. Time is also discretized such that

$$t_k = k\,\Delta t \qquad k = 0, \dots, N_t \tag{9}$$

For explicit time integration, the time step is governed by the CFL condition, $\Delta t = \frac{h}{2c_s}$ in the first problem. In the second problem, this time step is reduced to $\frac{h}{16c_s}$ to avoid oscillations and instabilities.

Here, $i$ and $j$ represent node numbers in the $x$ and $y$ directions, respectively, in the main FD grid and $k$ controls discretization in time.

The simulation starts with an initial prescribed shear stress distribution on the fault and a zero-displacement field everywhere in the medium. To avoid confusion, discretized field variable values updated with FD are represented with subscripts $i$ and $j$ while the ones on the strip boundary that are obtained by the SBI method are marked with subscript $l$. Suppose that the discretized values of displacement $u_l(t)$ and $u_{i,j}^k$, velocity $V_l(t)$, and shear stress $\tau_l(t)$ are known at time $t$. We will show how these values are updated in one evolution time step.

The updating scheme used for the boundary integral method here is based on the work of Lapusta et al. (2000) with some modifications to allow for the use of an independent spectral formulation.

1. Make first predictions for the values of displacement on the boundary at time $t + \Delta t$, based on known values at time $t$, as

$$u_l^{\pm^*}(t + \Delta t) = u_l^\pm(t) + V_l^\pm(t)\,\Delta t \tag{10}$$

2. Make a corresponding first prediction $f_l^{\pm^*}(t + \Delta t)$ of the functionals, using displacement predictions from step 1 and considering the slip rates to be constant through the time step $\Delta t$ and equal to $V_l(t)$. To do this, we first compute the Fourier coefficients of $V_l^\pm(t)$ and $u_l^{\pm^*}(t + \Delta t)$. To represent FFT operations, we use the following relations.

$$\dot{D}_n^\pm(t) = \sum_{l=1}^{N_x} V_l^\pm(t)\,\Omega_{N_x}^{(l-1)(n-1)} \tag{11}$$

$$D_n^{\pm^*}(t + \Delta t) = D_n^\pm(t) + \Delta t\,\dot{D}_n^\pm(t) \tag{12}$$

$$
\begin{aligned}
F_n^{\pm^*}(t + \Delta t) = \mp\mu|k_n|\Bigg[ & D_n^{\pm^*}(t + \Delta t) \\
& - \int_{\Delta t}^{t+\Delta t} W(|k_n|c_s t')\dot{D}_n^\pm(t + \Delta t - t')dt' \\
& - \dot{D}_n^\pm(t)\int_0^{\Delta t} W(|k_n|c_s t')dt' \Bigg]
\end{aligned} \tag{13}
$$

$$f_l^{\pm^*}(t + \Delta t) = \frac{1}{N_x}\sum_{n=1}^{N_x} F_n^{\pm^*}(t + \Delta t)\,\Omega_{N_x}^{-(l-1)(n-1)} \tag{14}$$

Here, $k_n = \frac{2\pi n}{\lambda}$ is the wave number and $\Omega_{N_x} = e^{\frac{(-2\pi i)}{N_x}}$ where $i = \sqrt{-1}$ and $W$ is the convolution kernel for the velocity formulation in the 2-D antiplane case (Lapusta et al. 2000).

3. Use the displacement values obtained in step 1 to calculate the displacement values on the boundary at the FD grid points. Since calculations are based on the values at the cell centers in the boundary integral method, interpolation is required to obtain the displacement values at the FD nodes (Figure 2).

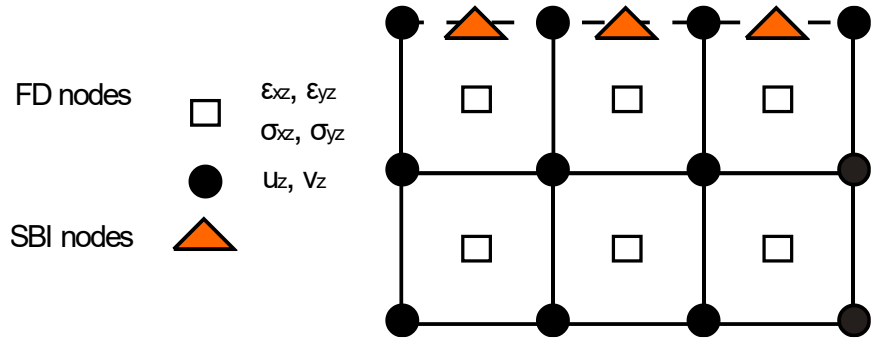

Figure 2- Configuration of the boundary nodes for a group of grid cells on the partly-staggered grid. The non-staggered grid has a similar arrangement except that all field variables in the FD method are defined at the corner nodes.

$$u_{i,N_y/2}^{k+1/2} = \frac{1}{2}\left(u_l^{+^*}(t+\Delta t) + u_{l+1}^{+}{}^*(t+\Delta t)\right) \quad l = 1, \dots, N_x - 1 \tag{15}$$

$$u_{i,-N_y/2}^{k+1} = \frac{1}{2}\left(u_l^{-^*}(t+\Delta t) + u_{l+1}^{-}{}^*(t+\Delta t)\right) \quad l = 1, \dots, N_x - 1 \tag{16}$$

Here, $i$ ranges from $-\frac{N_x}{2}+1$ to $\frac{N_x}{2}-1$.

At extreme nodes ($i = -\frac{N_x}{2}, \frac{N_x}{2}$), the displacements may be determined either by imposing periodic or classical absorbing boundary conditions to the lateral edges of the FD domain. In the examples to follow, the simulation time was chosen shorter than what is required for the waves to reach the lateral boundaries and reflect to interact with the solution and thus the results are insensitive to the particular choice of the lateral boundary conditions. We discuss this further in Section 4.

4. Solve the wave propagation problem for the interior points on the strip to get the discrete displacement values $u_{i,j}^{k+1}$ as well as the slip and traction values at the corresponding fault nodes.

5. Calculate the values of shear traction $\tau_l^*(t+\Delta t)$ at the SBI nodes on the boundary. In the non-staggered grid, this is done by using the displacement values calculated in steps 3 and 4 and a second-order accurate, one-sided FD scheme.

$$\tau_{i,N_y/2}^{k+1} = \frac{\mu}{2\Delta y}(3u_{i,N_y/2}^{k+1} - 4u_{i,N_y/2-1}^{k+1} + u_{i,N_y/2-2}^{k+1}) \tag{17}$$

$$\tau_{i,-N_y/2}^{k+1} = \frac{\mu}{2\Delta y}(-u_{i,-N_y/2+2}^{k+1} + 4u_{i,-N_y/2+1}^{k+1} - 3u_{i,-N_y/2}^{k+1}) \tag{18}$$

These values are then averaged to determine the traction at the SBI nodes.

$$\tau_l^{\pm^*}(t+\Delta t) = \frac{1}{2}(\tau_{i,\pm N_y/2}^{k+1} + \tau_{i+1,\pm N_y/2}^{k+1}) \qquad i = -\frac{N_x}{2}, \dots, \frac{N_x}{2}-1 \tag{19}$$

Here, $l$ ranges from 1 to $N_x$.

In the partly-staggered grid, however, for each SBI node on the interface, a unique quadratic function $\tau^{app}(y)$ is defined based on the discrete values of the shear stresses at the three neighboring FD nodes $\left(\tau_{i+\frac{1}{2},\frac{\pm N_y}{2}\mp\frac{1}{2}}^{k+1}, \tau_{i+\frac{1}{2},\frac{\pm N_y}{2}\mp\frac{3}{2}}^{k+1}, \tau_{i+\frac{1}{2},\frac{\pm N_y}{2}\mp\frac{5}{2}}^{k+1}\right)$ for $i$ ranging between $-\frac{N_x}{2}$ to $\frac{N_x}{2}-1$). The shear stress at the SBI node on the boundary is then extrapolated using this function.

6. Make first predictions of the values of velocities at time $t+\Delta t$ using the following formula.

$$\tau_l^{\pm^*}(t+\Delta t) = \tau_{0_l}^{\pm} \mp \frac{\mu}{c_s} V_l^{\pm^*}(t+\Delta t) + f_l^{\pm^*}(t+\Delta t) \tag{20}$$

7. Calculate the second prediction for displacement at time $t+\Delta t$ by

$$u_l^{\pm^{**}}(t + \Delta t) = u_l^{\pm}(t) + \frac{\Delta t}{2}\left[V_l^{\pm}(t) + V_l^{\pm^*}(t + \Delta t)\right]$$

(21)

8. Make a corresponding prediction $f_l^{\pm^{**}}(t + \Delta t)$ of the functional, using the $u_l^{\pm^{**}}(t + \Delta t)$ and assuming velocities equal to $\frac{1}{2}\left[V_l^{\pm}(t) + V_l^{\pm^*}(t + \Delta t)\right]$ throughout the time step. This assumption is consistent with step 7. The procedure is exactly as in step 2.

9. Use the corrected displacement values obtained in step 7 to calculate the displacement values at the FD nodes on the boundary and get a more accurate estimate of $\tau_l^{\pm^{**}}(t + \Delta t)$. The procedure is identical to that explained in steps 3 to 5.

10. Make final predictions $V_l^{\pm^{**}}(t + \Delta t)$ of the velocities as in step 6.

11. Set the values of the field quantities $u_l^{\pm}(t + \Delta t)$, $V_l^{\pm}(t + \Delta t)$ on the boundary equal to the predictions with the superscript double asterisks. Set the values of the field variables on the FD grid equal to the ones obtained in step 9.

12. Use the $u_{i,N_y/2}^{k+1}$ and $u_{i,-N_y/2}^{k+1}$ obtained in step 9 as the boundary conditions for the FD scheme in the next time step. Finally, return to step 1 to advance the solution forward.

## 3. Results

### 3.1. Slip-weakening crack with a low-velocity fault zone

We consider an 8-kilometer fault embedded in a heterogeneous, linear elastic medium with a density of 3334 kg/m³ as shown in Figure 3. The reduction in the shear velocity in the low-velocity zones (LVZ) may range from a few percent to as much as 60 percent (Huang & Ampuero 2011). Here we have chosen a mild velocity contrast (~5%) to demonstrate the adequacy of the hybrid scheme and we defer exploration of more extreme contrast values (Ma & Elbanna 2015; Huang & Ampuero 2012; Huang et al. 2014; Huang et al. 2016) to future work.

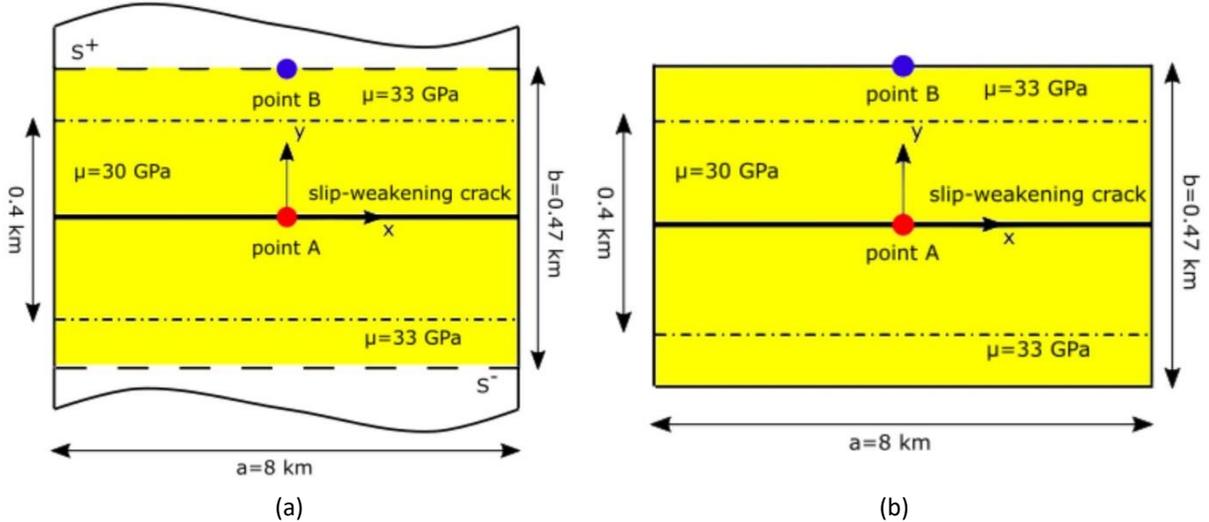

Figure 3- Model setup for antiplane fault embedded in a low velocity zone. (a) Configuration of the problem in the hybrid method. (b) Configuration of the problem in the Finite Difference method.

The host rock is assumed to have a shear modulus of 33 GPa. The shear modulus of the low velocity layer is 30 GPa and its width is 0.4 km. A slip-weakening friction law is used with a $D_c$ value equal to 0.4 meters. The friction coefficients and the initial stress distribution are as reported in Table 1. We have used the full SBI formulation with no mode truncation. In the hybrid method, we introduce virtual boundaries at 0.235 km on each side of the fault plane.

In principle, it is possible to take the virtual boundaries right at the boundary of the low velocity zone. We have not found our results to depend on the distance of the virtual boundary from the fault as long as the virtual strip fully includes the heterogeneity.

Figure 4 shows the spatial distribution of slip on the fault and displacement on the top strip boundary as computed from the two modeling approaches: the hybrid model and a pure finite difference model. The two methods show excellent agreement and no sign of artificial reflection from introducing the virtual boundary in the hybrid approach. Figure 5 shows the evolution of slip rate on the fault and the velocity on the top virtual strip's boundary. The agreement between the two methods in these plots is yet a stronger proof for their conformity. Figure 6 shows the time history plots for a point at the middle of the fault (point A) and at the middle of the top strip boundary (point B) as well as error plots for the same points showing convergence with mesh refinement. Once again, the two methods show very low levels of error. This is of high significance as it attests to the accuracy of the hybrid method and the exactness of the boundary condition introduced at the virtual boundaries. Figure 7 is provided to show and compare the resolution of the process zone for the three different meshes. It can be seen from these plots that the process zone narrows and converges with finer resolution. It should also be noted that the rupture speed is almost identical in the three resolutions and that the difference between the two highest resolutions is very small suggesting excellent convergence.

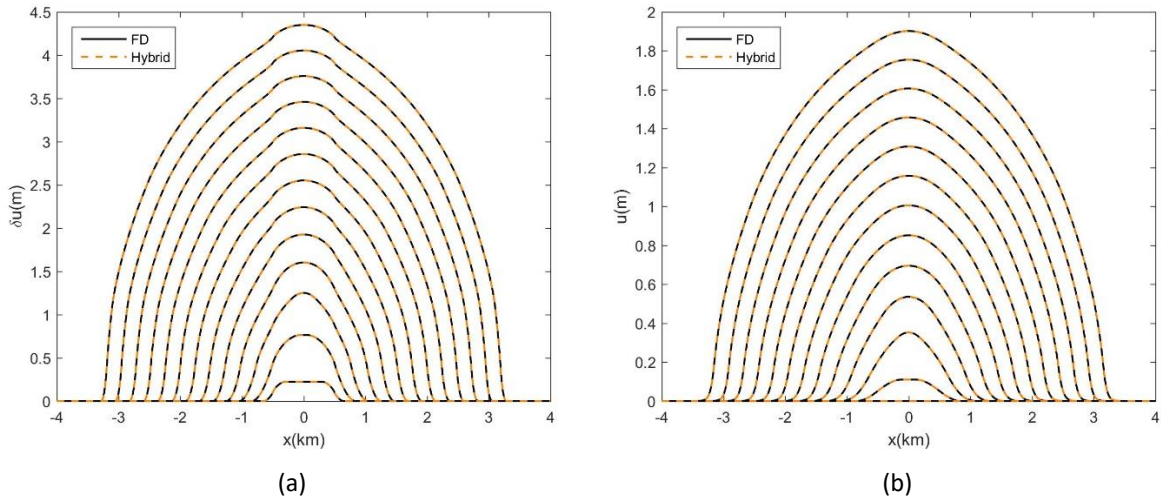

Figure 4- Comparison of solutions obtained from the finite difference method and the hybrid approach for the most refined mesh. (a) Evolution of slip on the fault plane every 77.6 milliseconds. (b) Displacement along the virtual boundary S⁺ every 77.6 milliseconds.

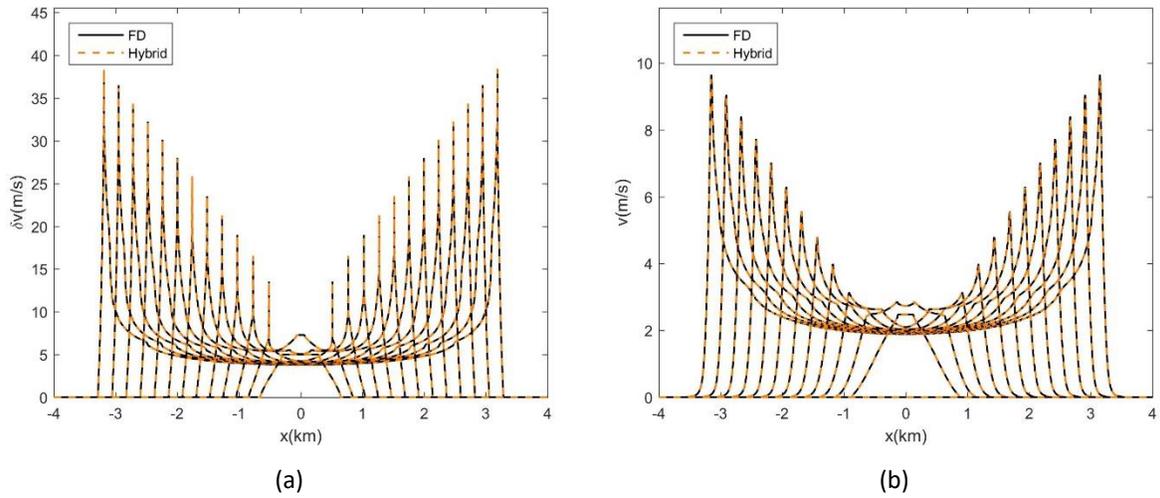

(a)　　　　　　　　　　　　　　　(b)

Figure 5- Comparison of the solutions obtained from the finite difference method and the hybrid approach for the most refined mesh. (a) Evolution of slip rate on the fault plane every 77.6 milliseconds. (b) Velocity along the virtual boundary S$^+$ every 77.6 milliseconds.

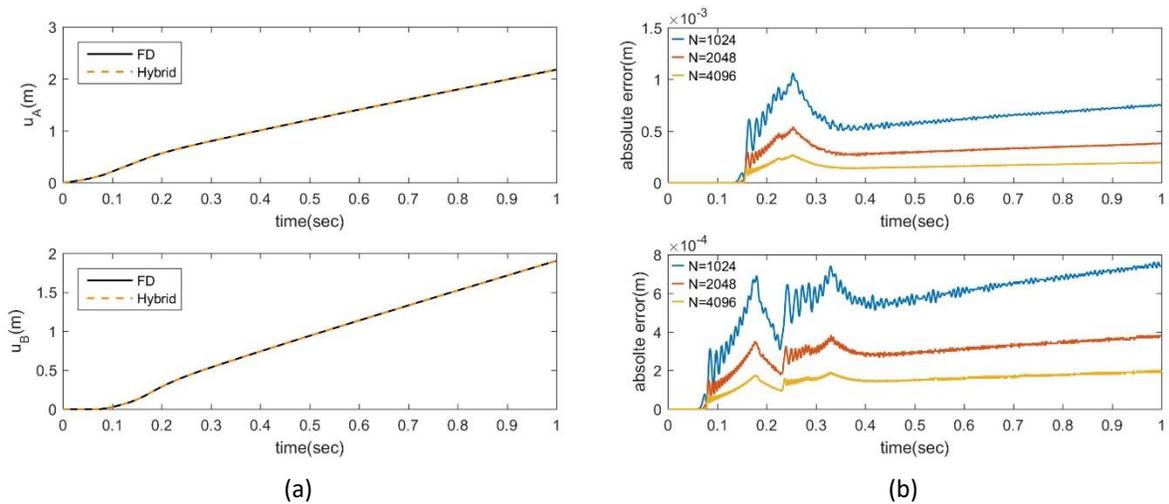

(a)　　　　　　　　　　　　　　　(b)

Figure 6- Evaluating the accuracy of the hybrid method in capturing the deformation history. (a) Time history for displacement at point A and displacement at point B. (b) Convergence with mesh refinement: the difference between the two methods further decreases with increasing mesh resolution.

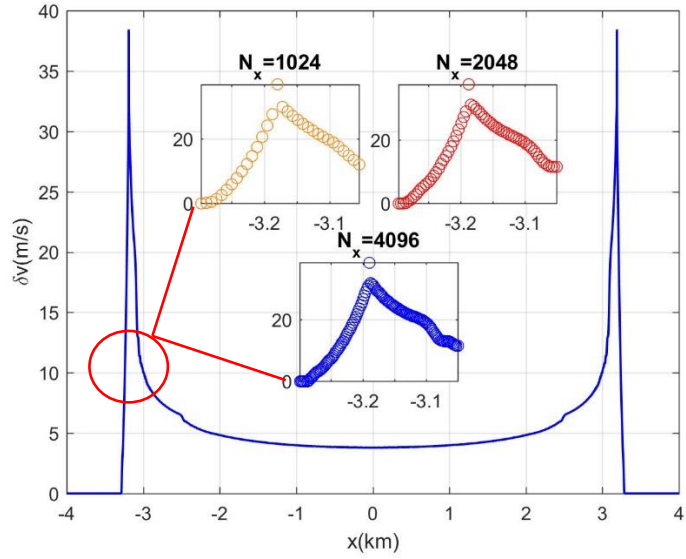

Figure 7- Resolution of the process zone for the three different meshes.

### 3.2. Slip-Weakening Crack with Plasticity

The setup of the problem is shown in Figure 8. The stress and friction parameters are the same as in the previous problem. The elastic properties are the same everywhere but we consider the possibility of co-seismic inelastic strain generation. A $J_2$ plasticity model is used with a radial return algorithm for integrating the stress update (Borja 2013). The yield stress is defined as $\tau_y = \tan(\phi)\,\sigma$ where $\sigma$ is the normal stress and $\phi$ is the effective friction angle. Assuming $\phi = 36°$ and $\sigma = 100\ MPa$ yields $\tau_y = 72\ MPa$. The flow rule is associative $\dot{\epsilon}^p = \dot{\lambda}\frac{\partial f}{\partial \tau}$ where $f$ is the $J_2$ yield function defined as $\sqrt{3\left(\tau_{zx}^2 + \tau_{zy}^2\right)} - \sqrt{3}\tau_y$ and $\dot{\lambda}$ is the plastic multiplier.

Other plasticity models, such as Mohr-Coulomb or Drucker-Prager, (Templeton & Rice 2008; Dunham et al. 2011), may be used. However, this will make little difference for our particular antiplane shear problem since the normal stress remains symmetric across the two sides of the fault. We deter the application of these popular models as well as other more sophisticated physics-based plasticity models (Ma & Elbanna 2017) to future work. Artificial viscosity and hourglass control are introduced in the model to suppress zero energy modes and numerical oscillations. The values of the viscosity and hourglass damping ($\boldsymbol{\eta}$ and $\boldsymbol{\chi}$) used are 1 and 0.7, respectively. The hourglass stiffness parameter ($\overline{\mathbf{Y}}$) is 1418 MPa.

Unlike in the previous case, we do not know a priori where to place the virtual boundaries. A reasonable initial guess is to estimate the thickness of the virtual strip to be of the order of 1/10 of the expected propagation distance. This is motivated by results from previous simulations of dynamic ruptures with off-fault plasticity (e.g. Templeton & Rice 2008; Dunham et al. 2011) which suggest that the region of inelastic strain grows only weakly with propagation distance. Later we will discuss possible solutions if the inelastic domain extends to reach the virtual boundaries

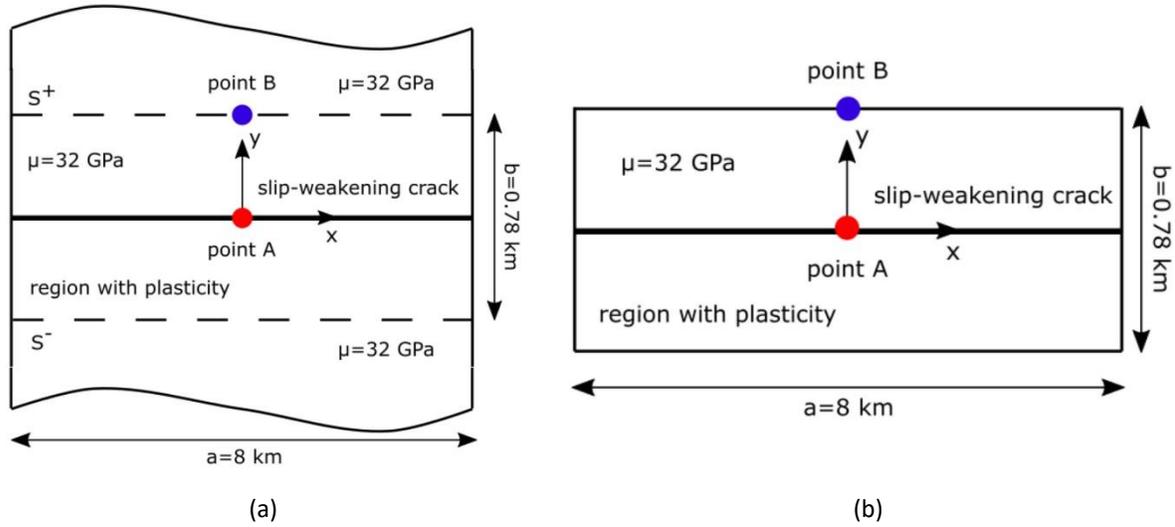

Figure 8- Model setup for antiplane fault with off-fault plasticity. (a) Configuration of the problem in the hybrid method. (b) Configuration of the problem in the Finite Difference method.

Figure 9 shows the spatial distribution of slip and displacement at different times for the hybrid and the FD methods and Figure 10 compares the solutions from both methods in terms of slip rate/ velocity. Figure 11 shows the time history plots for a point at the middle of the fault (point A) and at the middle of the boundary (point B) as well as error plots showing convergence with mesh refinement. Furthermore, we show in Figure 12 a comparison between the plastic strain distribution obtained from both methods at the end of the simulation. The plots show that the hybrid method's prediction matches that of Finite Difference. While both methods yield similar solutions on all three meshes, we have observed discrepancies in the velocity as well as plastic strain peaks among different meshes, when the same viscosity and hourglass damping parameters are used. This along with the strain localization effects that exist in the plastic strain plots have led us to believe that the damping parameters used are mildly mesh-dependent and should be calibrated based on mesh size. Figure 13 shows how using the same damping parameters in the three meshes affects the peak velocity in the cohesive zone. A separate study is likely required to look into the calibration of these mesh-dependent parameters for the anti-plane problem. Furthermore, the use of higher order schemes would eliminate the need for, at least, the hourglass regularization. For our purposes here, however, the fact that both methods produce the same results on each mesh is sufficient to show that the hybrid method is working as expected.

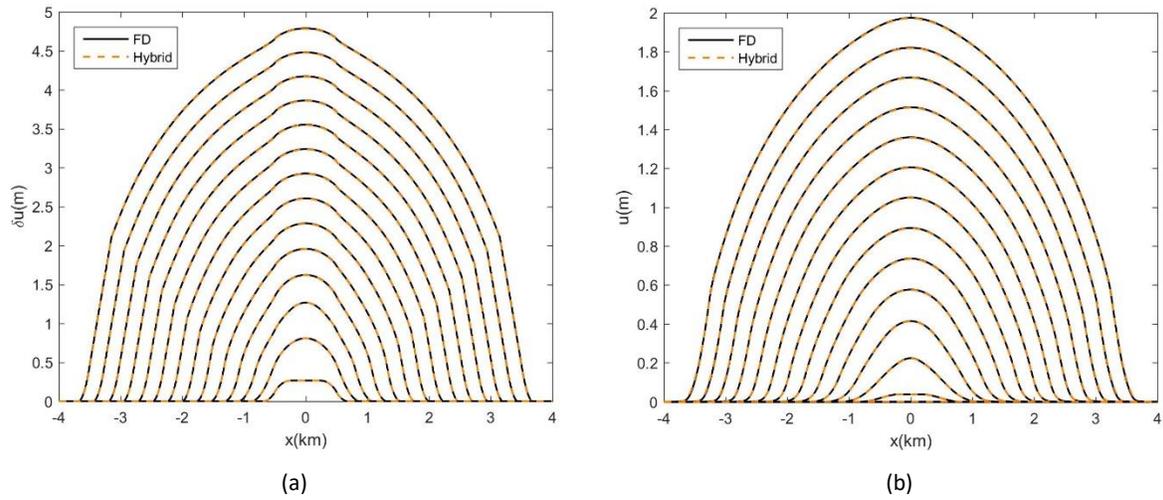

(a)                                     (b)

Figure 9- Comparison of solutions obtained from the finite difference method and the hybrid approach for the most refined mesh. (a) Evolution of slip on the fault plane every 70.5 milliseconds. (b) Displacement along the virtual boundary S⁺ every 70.5 milliseconds.

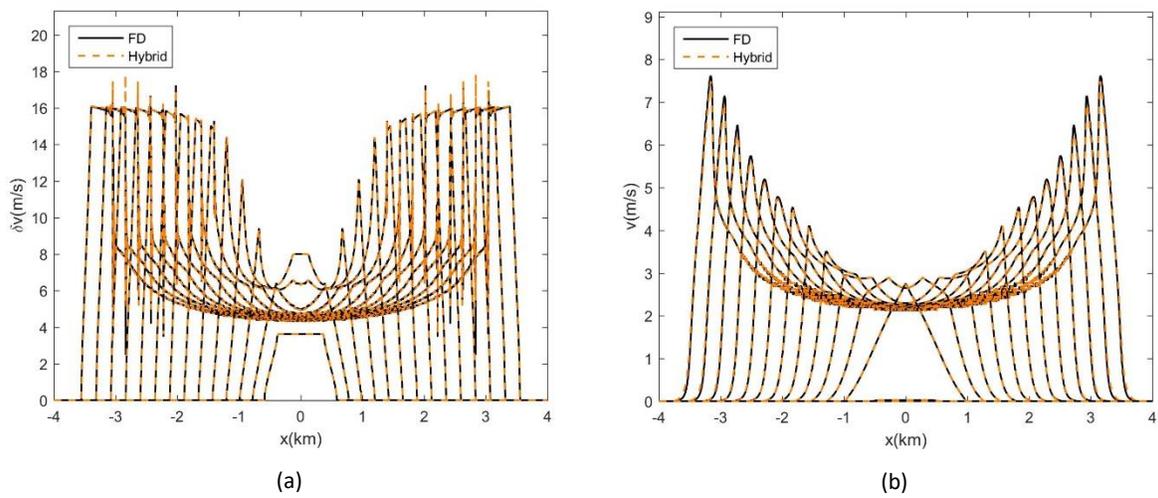

(a)                                     (b)

Figure 10- Comparison of solutions obtained from the finite difference method and the hybrid approach for the most refined mesh. (a) Evolution of slip rate on the fault plane every 70.5 milliseconds. (b) Velocity along the virtual boundary S⁺ every 70.5 milliseconds.

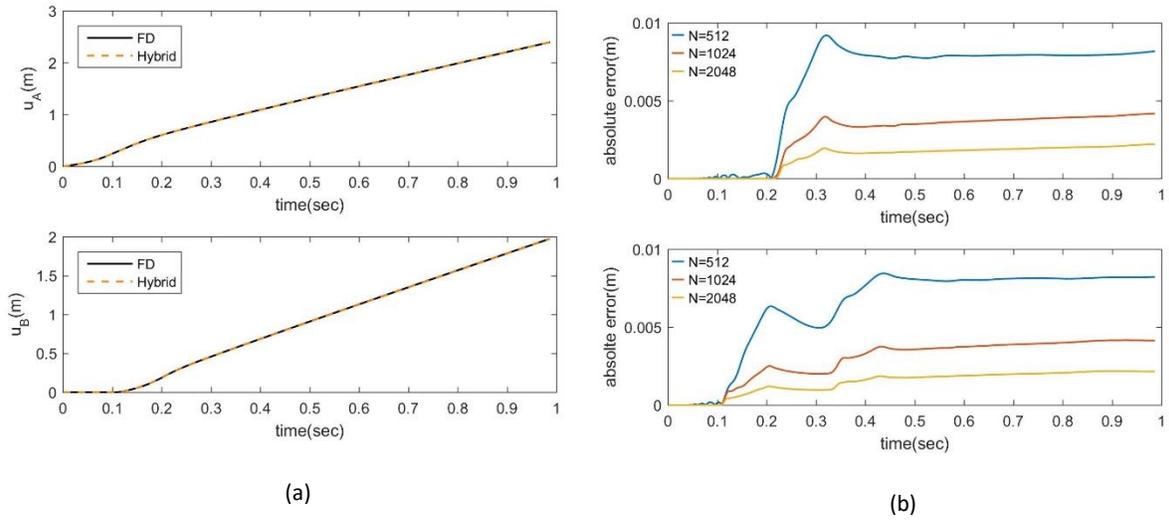

(a)                                                                      (b)

Figure 11- Evaluating the accuracy of the hybrid method in capturing the deformation history. (a) Time history for displacement at point A and displacement at point B (bottom). (b) Convergence with mesh refinement.

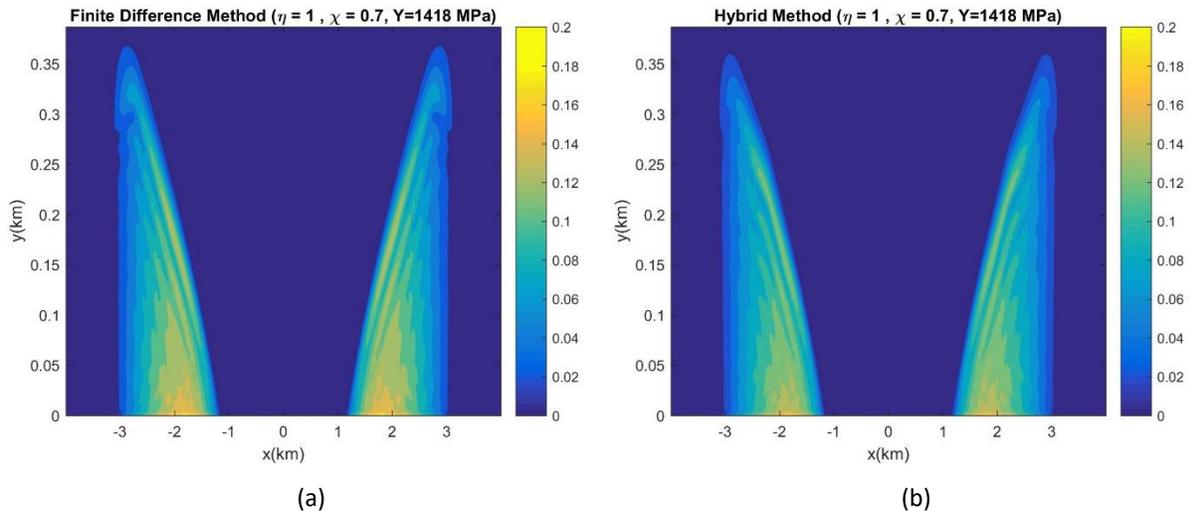

(a)                                                                      (b)

Figure 12- Spatial distribution of the plastic strain for the finest mesh in (a) the finite difference method and (b) the hybrid method.

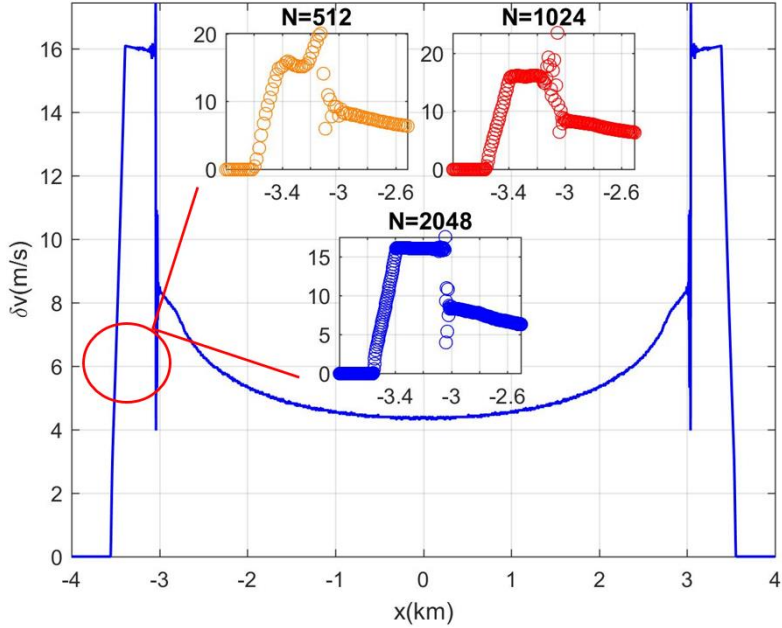

Figure 13- Resolution of the process zone for the three different meshes. $\eta = 1$ and $\chi = 0.7$.

## 4. Discussion

Although there has been significant progress in resolving the spatiotemporal complexities of earthquake ruptures, there's still a need to develop a numerical algorithm that is capable of long-time simulation of earthquake cycles in a bulk that may have material heterogeneity, material nonlinearity or fault surface complexity. Here, we have shown initial progress in developing a new hybrid numerical scheme in which the finite difference and the boundary integral equation methods are combined. This new method benefits from the flexibility of FD in handling heterogeneities and nonlinearities as well as the computational efficiency of SBI in modeling wave propagation without discretizing the full bulk. The saving in bulk discretization cost may enable simulating fault zone nonlinearities with unprecedented resolution and may allow simulating nonlinear problems on larger domains with the same computational cost.

We have shown that the hybrid method yields the same results as the full FD simulation and that it converges upon refinement. There is no signature for artificial wave reflections at the virtual boundaries which attest to the accuracy of the boundary conditions. In addition to the models shown in this paper, we have also tested the method in other setups including wave propagation from Gaussian and square sources as well as for rupture propagation on faults with different friction laws (Hajarolasvadi 2016). In all these cases, the method has shown excellent performance and demonstrated convergence with mesh refinement.

By limiting the spatial discretization to a small region near the fault plane encapsulating heterogeneities or bulk nonlinearities, more computational resources may be directed to explore additional complex physics within the fault zone. For example, it should be possible to simulate dynamics of rough faults at higher resolution than what is currently possible, incorporate gouge Multiphysics, and explicitly represent small-scale heterogeneities such as minor parallel faults and shear bands (Ma & Elbanna 2017).

The excellent performance of the hybrid scheme and the absence of artificial reflections from the virtual boundaries suggest that the method may be also used as an accurate near-field wave truncation algorithm. Current widely-used absorbing boundary conditions and layers must be taken sufficiently far from fault plane to avoid wave artifacts from interaction with domain boundaries. The spectral boundary integral equation, however, provides an exact boundary condition for all wave angles incidence. In that respect, our method is very close in spirit to the Dirichlet-to-Neuman maps in the pioneering works of Dan Givoli and collaborators (Harari, Patlashenko & Givoli 1998 and references therein). A point of departure in the current work is that we solve the non-local boundary condition in the Fourier domain benefiting from the planarity of the virtual boundaries. Convolutions in the real space correspond to multiplication of the Fourier components in the spectral domain and this locality of calculations is expected to significantly improve the computational costs and enable efficient parallelization.

There have been notable previous attempts for coupling the boundary integral method with bulk discretization methods including the finite element method (Bielak & MacCamy 1991). All the attempts we are aware of adopt a discretization of the spatial convolutions within the boundary integral in the real space. This usually leads to a densely-populated stiffness matrix for the coupled problem. One possible advantage of our proposed algorithm is to solve the convolution in the Fourier domain transforming the non-local boundary condition in space to a local one in the wave number. The complexity of the calculations in 2D, thus, reduces from $N^2$ to $NlogN$ (Lapusta et al. 2000).

One potential promising application of the proposed method is in the field of cycle simulations of earthquakes in a bulk containing near-field material heterogeneities, material nonlinearities or fault surface complexity. This may be the case for two reasons. First, the independent spectral boundary integral formulation enables accurate near-field truncation of the wave field eradicating the need for discretizing a significant portion of the bulk and thus leading to significant computational saving. Furthermore, unlike most other known absorbing boundary conditions or layers, the integral formulation is accurate in both dynamic and quasidynamic limits and thus will be capable of handling both the dynamic and interseismic portions of the seismic cycle. Second, adopting the spectral formulation will enable us to leverage the infrastructure developed by Lapusta et al. (2000) regarding mode truncation and adaptive time-stepping. For efficient simulation of the long term seismic history, an implicit time integration scheme will probably be required for the bulk method within the virtual strip. This is the focus of our ongoing work.

In this paper, we have considered an anti-plane shear rupture in an infinite bulk. More realistic rupture scenarios should include the free surface and eventually incorporate in-plane and 3D rupture geometries. The inclusion of the free surface is straightforward using the method of mirrors to enforce the stress-free boundary condition and modify the boundary integral formulation accordingly (Lapusta et al. 2000). The extension to in-plane rupture propagation is also feasible as the convolution kernels for the independent spectral formulation in this case are readily available (Geubelle & Breitenfeld 1997). For in-plane problems, however, the coupling between SBI and FD must be done for both the normal and shear stress components. With both the anti-plane and in-plane coupling available, the method may be directly extended to 3D.

Our current investigation has focused primarily on evaluating the accuracy of the hybrid method in calculating the solution of the field variables in the virtual strip and the crack surface without artificially reflecting waves off the computational strip's fault-parallel boundaries. This is usually the most challenging task for imposing near-field absorbing boundary conditions as the wave incidence angle strongly deviates from 90 degrees. Due to the large aspect ratio of the computational strip in fault rupture problems, the lateral boundaries of the computational strip will experience wave incidence at nearly 90 degrees. In this limit, almost all absorbing boundary conditions, including viscous dampers and indeed perfectly matching layers, will perform very well. We thus suggest that the lateral boundaries of the finite difference strip be modeled using any classical absorbing boundary conditions.

While the method is showing premise in a variety of applications, we acknowledge that it has some limitations. For example, we have assumed that the bulk away from the fault zone is homogeneous and linear elastic. This enabled us to directly employ the boundary integral formulation at the edges of the virtual strip. However, heterogeneities and anelasticity may exist in the far field as well. In this case, the application of the boundary integral formulation is no longer exact. Nonetheless, we hypothesize that rupture dynamics is most sensitive to local fault conditions and small-scale heterogeneities in the fault zone since these will directly affect the rupture tip and the process zone over which steep gradients in stress and particle velocity exist. The effect of a heterogeneity also depends on how far it is and how strong the contrast in its properties are from the bulk properties. The further the heterogeneity, the smaller its effect on the dynamics. For calculations of ground motion, however, this may be a critical issue since wave amplitude and phase at a site depends on the wave path. If the focus of an application is on the ground motion, we propose that the hybrid method may be used as a simulator to predict the source characteristics including fault plane slip and slip rate distribution. This information may then be fed back into a wave simulation code to track wave propagation in globally heterogeneous and inelastic media and to predict ground motion.

Another possible limitation is that for dynamic heterogeneities, such as off-fault plasticity, we may not know beforehand the extent to which the nonlinearities are expected to grow. This will require making an educated guess or implement a trial and error approach. For off-fault plasticity, we may take advantage of previous studies that suggest inelasticity or damage to extend to only a narrow region near the fault plane (Poliakov et al. 2002; Rice et al. 2005; Templeton & Rice 2008; Dunham et al. 2011). This is the approach we have adopted so far. It is possible, however, to adopt an adaptive scheme. For example, if the inelastic region is found to approach the virtual boundaries we may push the boundaries away from the fault plane and restart the calculations from the last time step. For that purpose, we will need the solution for displacement and velocity in the new strip added by expanding the original virtual domain. This may be directly calculated using the history of the solution on the original virtual boundary and a direct application of the representation theorem.

## 5. Conclusion

A new hybrid scheme has been developed to enable simulation of wave propagation problems in unbounded domains with near source heterogeneities. The method is primarily suited for earthquake rupture simulations where the wave source is extending predominantly in one direction (or plane). The main conclusions of the paper are summarized as follows:

- The hybrid method yields similar results as the full FD simulation at a fraction of discretization cost and the method converges upon refinement.

- The hybrid method has potential for being used in the field of cycle simulations of earthquake in a medium with heterogeneities and/or nonlinearities since FD will enable capturing these features confined in the virtual strip without the need to discretize the whole bulk while the spectral boundary integral will enable mode truncation and adaptive time-stepping to resolve the various scales in time (e.g. both rapid slip and interseismic deformations).

- The hybrid method can also be potentially used for studying additional small-scale physics within the fault zone by saving memory and other resources that would otherwise have to be consumed by a full volume-discretization based numerical schemes.

- The excellent performance of the hybrid scheme and the absence of artificial reflections from the virtual boundaries suggest that the method may be also used as an accurate near-field wave truncation algorithm.

## Appendix A: Finite Difference Equations

### A.1.  Traction-at-split-node (TSN) Method

The traction-at-split node method presented here is adopted from Moczo et al. (2007).

Consider two half-spaces $S^+$ and $S^-$ discretized by two FD grids and two arbitrary grid points $p.n.^+$ and $p.n.^-$ on each side. Define the outer normal $\vec{n}$ to the surface $S^-$ pointing towards the surface $S^+$.

The acceleration for these nodes can be expressed as $\vec{\ddot{u}}^\pm = \frac{\vec{R}^\pm}{M^\pm}$ where $R^\pm$ are forces due to deformation in each halfspace and $M^\pm$ represent the masses associated with each node. We couple the surfaces to simulate a fault by defining a contact force $\overline{T^c}(\vec{n})$ between the two. The expression for the acceleration of these partial nodes is then given by:

$$\vec{\ddot{u}}^\pm = \frac{\vec{R}^\pm \mp A\left[\overline{T^c}(\vec{n}) - \vec{T_0}\right]}{M^\pm} \tag{A1}$$

where $A$ represents the area of the fault surface associated with each partial node and $T_0$ is the traction at equilibrium state.

Note that for the 2-D antiplane problem at hand, these equations reduce to

$$\ddot{u}_z^{\,\pm}(t) = \frac{R_z^\pm(t) \mp A[T^c(t) - T_0]}{M^\pm} \tag{A2}$$

We determine the traction between the two surfaces at time $t$ by finding a constraint traction $T^{ct}(t)$ that assures zero slip-rate before the two nodes start slipping and when the slipping ceases. To do so, the velocities $\dot{u}_z^\pm\left(t + \frac{dt}{2}\right)$ and displacements $u_z^\pm(t + dt)$ of the partial nodes are approximated by a central difference scheme in time.

$$\dot{u}_z^\pm\left(t + \frac{dt}{2}\right) = \dot{u}_z^\pm\left(t - \frac{dt}{2}\right) + dt\,\ddot{u}_z^{\,\pm}(t) \tag{A3}$$

$$u_z^\pm(t + dt) = u_z^\pm(t) + dt \, \dot{u}_z^\pm(t + \frac{dt}{2}) \tag{A4}$$

Therefore, the slip rate is expressed as

$$
\begin{aligned}
dV\left(t + \frac{dt}{2}\right) = dV\left(t - \frac{dt}{2}\right) \\
+ dt \, B \left\{ \frac{M^- R^+(t) - M^+ R^-(t)}{A.(M^- + M^+)} - [T^c(t) - T^0] \right\}
\end{aligned}
\tag{A5}
$$

with $B = \frac{A(M^+ + M^-)}{M^+ M^-}$.

To find the constraint traction, we require that $dV\left(t + \frac{dt}{2}\right) = 0$. Therefore,

$$T^{ct}(t) = T^0 + \frac{dt^{-1} M^- M^+ dV\left(t - \frac{dt}{2}\right) + M^- R^+(t) - M^+ R^-(t)}{A.(M^- + M^+)} \tag{A6}$$

$T^c(t)$ now can be determined as follows

$$
\begin{aligned}
\text{If } T^{ct}(t) \leq S \text{ then } T^c(t) = T^{ct}(t) \\
\text{If } T^{ct}(t) > S \text{ then } T^c(t) = S
\end{aligned}
\tag{A7}
$$

where $S$ is the frictional strength. These expressions illustrate that if the constraint traction is less than the frictional strength, then the two nodes cannot slip. Therefore, the traction on the fault would be equal to the constraint traction. However, when the value of $T^{ct}(t)$ exceeds the frictional strength, slip starts accumulating and the traction between the two surfaces must be equal to the frictional strength.

To compute the forces R we adopt the finite difference discretization of Day et al. (2005), tailored for the antiplane problem, as briefly discussed below.

## A.2.  Plasticity Implementation

A partly-staggered finite difference grid is defined where the displacement and velocity field values are defined at the corner nodes $Q$ while the stress and strain field values are defined at the central nodes $\bar{Q}$ (Figure A1).

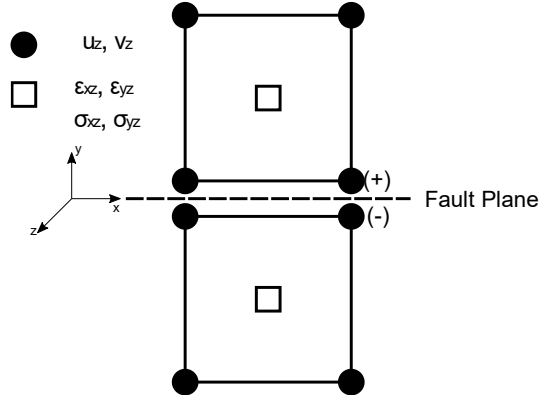

Figure A1- Split-node geometry of the FD method for two square cells.

The strain tensor $\boldsymbol{\epsilon} = \frac{1}{2}(\nabla \boldsymbol{u} + \nabla \boldsymbol{u}^T)$ for an arbitrary node $\bar{Q}$, $\boldsymbol{\epsilon}^{\bar{Q}}(t)$ can be expressed in terms of the discretized values of displacements at the corner nodes, $\boldsymbol{u}^Q(t)$. For the antiplane problem, the only nonzero component of displacement is $u_z(x, y, t)$. Therefore, the nonzero strain components are $\epsilon_{xz}$ and $\epsilon_{yz}$. Using central difference in time,

$$\dot{\boldsymbol{\epsilon}}^{\bar{Q}}\left(t - \frac{dt}{2}\right) = \frac{\boldsymbol{\epsilon}^{\bar{Q}}(t) - \boldsymbol{\epsilon}^{\bar{Q}}(t - dt)}{dt} \tag{A8}$$

Using a predictor-correcter approach, it is first assumed that the total strain rate is fully elastic. Stresses are then calculated as

$$\boldsymbol{\tau}_{elp}^{\bar{Q}}(t) = \boldsymbol{\tau}^{\bar{Q}}(t - dt) + 2\mu \dot{\boldsymbol{\epsilon}}^{\bar{Q}}\left(t - \frac{dt}{2}\right)dt \tag{A9}$$

at the cell centers and the yield criteria is checked to determine the veracity of such assumption. Here, $\boldsymbol{\tau}_{elp}^{\bar{Q}}(t)$ is the elastic predictor for the stress at time $t$. If $\sqrt{\left(\tau_{xz}^2 + \tau_{yz}^2\right)} - \tau_y \le 0$ then the assumption holds and $\boldsymbol{\tau}^{\bar{Q}}(t) = \boldsymbol{\tau}_{elp}^{\bar{Q}}(t)$. Otherwise, plasticity is active. To proceed after the onset of yield, we assume an additive decomposition of the strain rate as follows:

$$\dot{\boldsymbol{\epsilon}}^{\bar{Q}}\left(t - \frac{dt}{2}\right) = \dot{\boldsymbol{\epsilon}}_{el}^{\bar{Q}}\left(t - \frac{dt}{2}\right) + \dot{\boldsymbol{\epsilon}}_{pl}^{\bar{Q}}\left(t - \frac{dt}{2}\right) \tag{A10}$$

where $\dot{\boldsymbol{\epsilon}}_{el}^{\bar{Q}}$ and $\dot{\boldsymbol{\epsilon}}_{pl}^{\bar{Q}}$ are the elastic and plastic components of the strain rate, respectively. A radial return algorithm (Borja 2013) is used to bring the predictor stress that has overshot back to the yield surface. The rate equations used are

$$\dot{\boldsymbol{\tau}}^{\bar{Q}} = 2\mu\left(\dot{\boldsymbol{\epsilon}}^{\bar{Q}} - \dot{\boldsymbol{\epsilon}}_{pl}^{\bar{Q}}\right), \qquad \dot{\boldsymbol{\epsilon}}_{pl}^{\bar{Q}} = \dot{\lambda}\frac{\partial f}{\partial \boldsymbol{\tau}^{\bar{Q}}} = \dot{\lambda}\hat{\boldsymbol{n}} \tag{A11}$$

where $f$ is the yield function and $\hat{\boldsymbol{n}}$ is the unit direction of the strain rate. The return mapping procedure dictates that the updated shear tensor at time $t$ must have the same direction as the elastic predictor but its norm is scaled to the radius of the yield surface. Therefore, the plastic corrector just brings the elastic predictor stress back to the yield surface. This enables us to write the shear stress as

$$\boldsymbol{\tau}^{\bar{Q}}(t) = \frac{\tau_y}{|\boldsymbol{\tau}_{elp}^{\bar{Q}}(t)|}\boldsymbol{\tau}_{elp}^{\bar{Q}}(t) \tag{A12}$$

Using these new values, the elastic portion of the strain rate can be expressed as

$$\dot{\boldsymbol{\epsilon}}_{el}^{\bar{Q}}(\text{t} - \frac{dt}{2}) = \frac{\boldsymbol{\tau}^{\bar{Q}}(t) - \boldsymbol{\tau}^{\bar{Q}}(t - dt)}{2\mu dt} \tag{A13}$$

Consequently, $\dot{\boldsymbol{\epsilon}}_{pl}^{\bar{Q}}(\text{t} - \frac{dt}{2})$ can be determined by subtracting $\dot{\boldsymbol{\epsilon}}_{el}^{\bar{Q}}$ from the total strain rate. Furthermore,

$$\boldsymbol{\epsilon}_{pl}^{\bar{Q}}(\text{t}) = \boldsymbol{\epsilon}_{pl}^{\bar{Q}}(\text{t} - \text{dt}) + \dot{\boldsymbol{\epsilon}}_{pl}^{\bar{Q}}(\text{t} - \frac{dt}{2})dt \tag{A14}$$

To suppress high frequency numerical oscillations, artificial viscosity and hourglass control are included in the model. The introduction of the artificial viscosity, leads to the damping stress tensor $\bar{\sigma}$.

$$\overline{\boldsymbol{\sigma}}^{\bar{Q}}(t) = 2\,\eta\,\Delta t\,\mu\,\dot{\boldsymbol{\epsilon}}^{\bar{Q}}(t - \frac{dt}{2}) \tag{A15}$$

where $\eta$ is the damping constant.

Hourglass control is also included by adding stiffness and damping forces which stabilize the null modes. The implementation is briefly explained in section A.3.

The total nodal force $R_z^Q(t)$ is therefore calculated by adding of $\tau^{\bar{Q}} - \tau_0^{\bar{Q}}$, $\bar{\sigma}^{\bar{Q}}$ and resistance forces due to the hourglass modes. It follows that for the interior nodes, $\ddot{u}_z{}^Q(t) = \frac{R_z^Q(t)}{M^Q}$ where $M^Q$ is the average mass associated with each node. Finally, the discretized values of the velocities and displacements are obtained using a central difference scheme in time.

$$\dot{u}_z^Q\left(t + \frac{dt}{2}\right) = \dot{u}_z^Q\left(t - \frac{dt}{2}\right) + dt\,\ddot{u}_z{}^Q(t) \tag{A16}$$

$$u_z^Q(t + dt) = u_z^Q(t) + dt\,\dot{u}_z^Q(t + \frac{dt}{2}) \tag{A17}$$

The same procedure may be repeated for the nodes on the fault plane to get the values of $R_z^{\pm Q}(t)$. Next, the value of $T^{ct\,Q}(t)$ is computed at each node on the fault using equation (A6). These values are compared with the frictional strength of the fault at each node $S^Q(t)$ to determine the traction values $T^{c\,Q}(t)$. From there, equations A3 and A4 are used to determine the values of slip rate $V^Q\left(t + \frac{dt}{2}\right)$ and slip $\delta u^Q(t + dt)$ on the fault.

## A.3.   Hourglass Control

We start by degrading the equations presented in Day et al. (2005) in 3D to 2D in a consistent manner. For the antiplane shear problem, the only nonzero component of $\boldsymbol{H}$ (the hourglass force) is $H_z$. The general coordinate system for the 3D formulation and the relevant hourglass mode for the antiplane problem are shown in Figure A2.

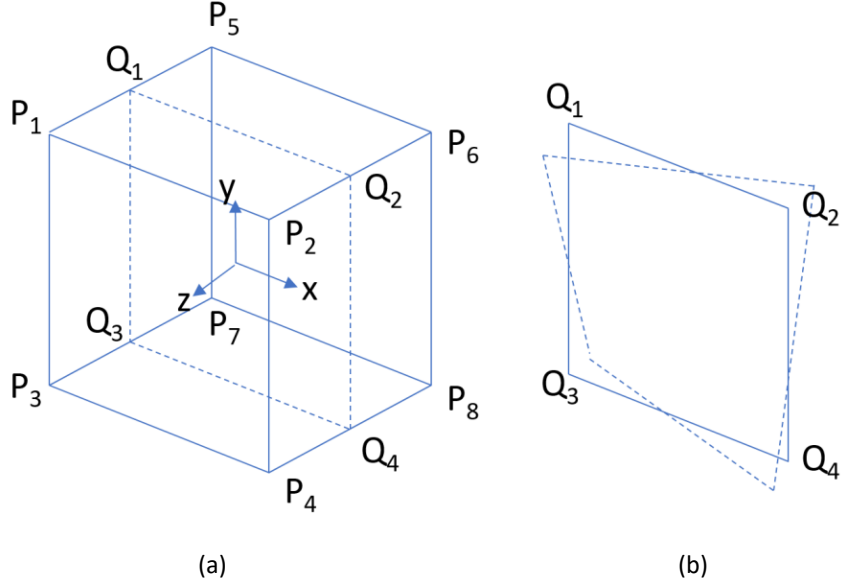

(a)                                     (b)

Figure A2-The hourglass instability. (a) A Schematic of a finite element cell in a general 3D simulation (after Day et al. 2005). For the antiplane shear problem, the relevant element is the dashed square ($Q_1 Q_2 Q_3 Q_4$) which is degraded from the 3D element by imposing the appropriate translational invariance along the z-coordinate. (b) The relevant hourglass mode for the antiplane shear.

The hourglass force is given by (Day et al. 2005)

$$H_{\bar{Q}}^i = \frac{A_D^i}{(\Delta z)_{\bar{Q}}^2} \left( u_{\bar{Q}+D} + \chi_{\bar{Q}} \Delta t \dot{u}_{\bar{Q}+D} \right) \tag{A18}$$

In this formula, the index D is defined as $D = [\frac{d_x}{2}, \frac{d_y}{2}, \frac{d_z}{2}]$ where $d_\nu$ with $\nu = x, y, z$ can be either 1 or -1. This index also implies a summation over all eight possible values of $D$ on the right hand side.

Moreover, $A_D^i$ is defined as

$$A_D^1 = d_x d_y d_z, \qquad A_D^2 = d_x d_y, \qquad A_D^3 = d_y d_z, \qquad A_D^4 = d_x d_z \tag{A19}$$

and $\chi_{\bar{Q}}$ is the hourglass viscosity.

For convenience, we introduce $W_{\bar{Q}+D} = u_{\bar{Q}+D} + \chi_{\bar{Q}} \Delta t \dot{u}_{\bar{Q}+D}$.

$$H_{\bar{Q}}^1 = \frac{1}{(\Delta_z)^2} \left( -W_{P_1} + W_{P_2} + W_{P_3} - W_{P_4} + W_{P_5} - W_{P_6} - W_{P_7} + W_{P_8} \right) \tag{A20}$$

$$H_{\bar{Q}}^2 = \frac{1}{(\Delta_z)^2} \left( -W_{P_1} + W_{P_2} + W_{P_3} - W_{P_4} - W_{P_5} + W_{P_6} + W_{P_7} - W_{P_8} \right) \tag{A21}$$

$$H_{\bar{Q}}^3 = \frac{1}{(\Delta_z)^2} \left( W_{P_1} + W_{P_2} - W_{P_3} - W_{P_4} - W_{P_5} - W_{P_6} + W_{P_7} + W_{P_8} \right) \tag{A22}$$

$$H_{\bar{Q}}^4 = \frac{1}{(\Delta_z)^2} \left( -W_{P_1} + W_{P_2} - W_{P_3} + W_{P_4} + W_{P_5} - W_{P_6} + W_{P_7} - W_{P_8} \right) \tag{A23}$$

However, in the antiplane shear problem, $u$ and $\dot{u}$ do not vary with $z$. Therefore,

$$W_{P_1} = W_{P_5}, \qquad W_{P_2} = W_{P_6}, \qquad W_{P_3} = W_{P_7}, \qquad W_{P_4} = W_{P_8} \tag{A24}$$

Imposing these equalities yields the following results.

$$H_{\hat{Q}}^1 = H_{\hat{Q}}^3 = H_{\hat{Q}}^4 = 0 \tag{A25}$$

$$H_{\hat{Q}}^2 = \frac{1}{(\Delta_z)^2} 2\left(-W_{Q_1} + W_{Q_2} + W_{Q_3} - W_{Q_4}\right) \tag{A26}$$

Using a similar approach Equation (A11) in Day et al., 2005 can be written in the following manner

$$F_Q = -2\bar{Y}\frac{(\Delta_x)^2}{\Delta z}\left(-H_{\hat{Q}_1}^2 + H_{\hat{Q}_2}^2 + H_{\hat{Q}_3}^2 - H_{\hat{Q}_4}^2\right) \tag{A27}$$

Where $\bar{Y}$ is the hourglass stiffness defined as $\frac{1}{12}\rho c_s^2(1 - \frac{c_s^2}{c_p^2})$ with $c_s$ and $c_p$ as the $P$ and $S$ wave speed. Therefore, for an arbitrary value of $\Delta z$ throughout the whole scheme, we can reduce these 3D equations to 2D equivalents.

## Appendix B: Independent Spectral Formulation

This appendix follows after Geubelle & Rice (1995) in its derivation of the independent SBI equations. Consider the problem setup shown in Figure 1. The only nonzero component of displacement, in this case, is $u_z$ and we have

$$c_s^2 \, u_{z,\alpha\alpha} = u_{z,tt} \qquad \alpha = x, y \tag{B1}$$

$$c_s^2 \, \Delta u_z = u_{z,tt} \tag{B2}$$

Let's examine one particular spectral component,

$$u_z(x, y, t) = e^{iqx}\, \Omega(y, t; q) \tag{B3}$$

The Laplace transform in the time domain is

$$\hat{f}(p) = L[f(t)] = \int_0^\infty e^{-pt} f(t) dt \tag{B4}$$

Rewriting Eq. (B2), we get

$$c_s^2 \left(\frac{\partial^2 u_z}{\partial x^2} + \frac{\partial^2 u_z}{\partial y^2}\right) = u_{z,tt} \tag{B5}$$

Applying the Fourier transform,

$$c_s^2 \left(-q^2\Omega + \frac{\partial^2 \Omega}{\partial y^2}\right) = \Omega_{tt} \tag{B6}$$

Applying the Laplace transform,

$$c_s^2 \left(-q^2\hat{\Omega} + \frac{\partial^2 \hat{\Omega}}{\partial y^2}\right) = p^2\hat{\Omega} \tag{B7}$$

Define $\widehat{\Omega}'' = \frac{\partial^2 \widehat{\Omega}}{\partial y^2}$. Eq. (B7) can be rewritten as a second-order ODE.

$$\widehat{\Omega}''(y,p;q) = q^2 \left(1 + \frac{p^2}{q^2 c_s^2}\right) \widehat{\Omega}(y,p;q) \tag{B8}$$

Introduce,

$$\alpha_s = \sqrt{1 + \frac{p^2}{q^2 c_s^2}} \tag{B9}$$

Eq. (B8) can be rewritten as,

$$\widehat{\Omega}''(y,p;q) = q^2 \alpha_s^2 \, \widehat{\Omega}(y,p;q) \tag{B10}$$

The PDE is now reduced to a second-order ODE that we can solve.

$$characteristic\ equation: \quad r^2 - q^2\alpha_s^2 = 0 \quad \rightarrow \quad r = \mp|q|\alpha_s \tag{B11}$$

Considering the radiation condition for the upper half-space and ignoring the unbounded solution, we will get

$$\widehat{\Omega}(y,p;q) = \widehat{\Omega}_0(p;q) e^{-|q|\alpha_s y} \tag{B12}$$

where $\widehat{\Omega}_0(p;q) = \widehat{\Omega}(0,p;q)$.

We are only concerned with the tractions along the fracture plane $y = 0$ and the resulting displacements. Therefore, define the Fourier coefficients by

$$u_z^+ (x, y = 0^+, t) = U_z^+(t;q)e^{iqx} \tag{B13}$$

Use Eq. (B3) and Eq. (B12) to get

$$\hat{u}_z(x,y,p) = e^{iqx}\widehat{\Omega}_0 e^{-|q|\alpha_s y}, \tag{B14}$$

$$\hat{u}_z(x,y=0^+;p) = \widehat{\Omega}_0 e^{iqx}, \tag{B15}$$

$$\hat{u}_z(x,y,p) = e^{iqx}\widehat{U}_z^+(p;q)e^{-|q|\alpha_s y} \tag{B16}$$

Writing $\tau_j(x,t)$ for the traction component of stress along the fracture plane

$$\tau_z^+(x,t) = \sigma_{yz}^+(x,y=0^+,t) = T_z^+(t;q)e^{iqx} \tag{B17}$$

We know,

$$\sigma_{ij} = \lambda\delta_{ij}u_{k,k} + \mu\big(u_{i,j} + u_{j,i}\big) \tag{B18}$$

where $\lambda$ and $\mu$ are Lamé constants.

$$\sigma_{yz} = \mu\big(u_{y,z} + u_{z,y}\big) = \mu\, u_{z,y} \tag{B19}$$

$$\hat{\sigma}_{yz} = \mu\,(\hat{u}_z)_{,y} = -\mu|q|\alpha_s e^{iqx}\widehat{U}_z(p;q)e^{-|q|\alpha_s y}, \tag{B20}$$

$$\hat{\tau}_z^+(x,p) = \hat{\sigma}_{yz}^+(x,y=0^+;p) = \widehat{T}_z^+(p;q)e^{iqx}, \tag{B21}$$

And,

$$\hat{\sigma}_{yz}^+(x,y=0^+,p) = -\mu|q|\alpha_s\widehat{U}_z(p;q)e^{iqx} \tag{B22}$$

Therefore,

$$\hat{T}_z^+(p;q) = -\mu|q|\alpha_s \hat{U}_z(p;q) \tag{B23}$$

The right-hand side of this equation can be rewritten by explicitly extracting the instantaneous response $-\frac{\mu p}{c_s} \hat{U}_z(p;q)$.

Hence,

$$\hat{T}_z^+(p;q) = -\frac{\mu p}{c_s} \hat{U}_z(p;q) - \mu|q| \left( \alpha_s - \frac{p}{|q|c_s} \right) \hat{U}_z(p;q) \tag{B24}$$

Back in the time domain, we have

$$\tau_z^+(x,t) = \tau_z^{0+}(x,t) - \frac{\mu}{c_s} \dot{u}_z^+(x,t) + f_z^+(x,t) \qquad upper-half\,space \tag{B25}$$

Following a similar procedure for the lower halfspace and imposing the radiation condition for $y < 0$, will similarly lead to

$$\tau_z^-(x,t) = \tau_z^{0-}(x,t) + \frac{\mu}{c_s} \dot{u}_z^-(x,t) + f_z^-(x,t) \qquad lower\,hal-space \tag{B26}$$

## Acknowledgements

The authors are grateful to Nadia Lapusta, Eric Dunham, J.-P Ampuero and Phillipe Geubelle for stimulating discussions. We also thank Alice-Agnes Gabriel and an anonymous reviewer for their constructive comments and suggestions that helped improve the paper. This work is supported by the Center for Geologic Storage of CO2, an Energy Frontier Research Center funded by the U.S. Department of Energy (DOE), Office of Science, Basic Energy Sciences (BES), under Award # DE-SC0012504.

## References

Ampuero, J.P., 2002. Etude physique et numérique de la nucléation des séismes, PhD thesis, Université Paris VII, Paris.

Ampuero, J.-P., 2004. Introduction to computational earthquake dynamics: a sample problem.

Ampuero, J.-P., 2008. SBIEMLAB, MATLAB code, http://web.gps.caltech.edu/~ampuero/software.html

Andrews, D. J., 1976. Rupture propagation with finite stress in antiplane strain, J Geophys. Res., 81(20), 3575. https://doi.org/10.1029/JB081i020p03575

Aochi, H., Ulrich, T., Ducellier, A., Dupros, F., & Michea, D., 2013. Finite difference simulations of seismic wave propagation for understanding earthquake physics and predicting ground motions: Advances and challenges, Journal of Physics: Conference Series, 454(1), 12010. https://doi.org/10.1088/1742-6596/454/1/012010

Archuleta, R. J., & Day, S. M., 1980. Dynamic Rupture in a Layered Medium: The 1966 Parkfield Earthquake, Bulletin of the Seismological Society of America, 70 (3). Seismological Society of America: 671–89.

Benjemaa, M., Glinsky, N., Cruz-Atienza, V. M., Virieux, J., Piperno, S., 2007. Dynamic non-planar crack rupture by a finite-volume method, Geophys. J. Int., 171,271–285, doi:10.1111/j.1365-246X.2006.03500.x.

Berenger, J.-P., 1994. A Perfectly Matched Layer for the Absorption of Electromagnetic Waves, Journal of Computational Physics. doi:10.1006/jcph.1994.1159.

Bettess, P., 1977. Infinite Elements, International Journal for Numerical Methods in Engineering, 11 (1). John Wiley & Sons, Ltd: 53–64. doi:10.1002/nme.1620110107.

Bielak, J., & MacCamy, R. C., 1991. Symmetric finite element and boundary integral coupling methods for fluid-solid interaction. Quart. Appl. Math., 49(1), 107–119.

Boore, D. M., Larner K. L., Aki K., 1971. Comparison of Two Independent Methods for the Solution of Wave-Scattering Problems: Response of a Sedimentary Basin to Vertically Incident SH Waves, Journal of Geophysical Research 76 (2): 558–69. doi:10.1029/JB076i002p00558.

Borja, R. I., Plasticity Modeling & Computation, pp 31--58, Springer-Verlag Berlin Heidelberg, https://doi.org/10.1007/978-3-642-38547-6.

Cochard, A., Madariaga, R., 1994. Dynamic faulting under rate-dependent friction. Pure and Applied Geophysics PAGEOPH, 142(3–4), 419–445. https://doi.org/10.1007/BF00876049.

Cruz-Atienza, V. M., Virieux, J., Aochi, H., 2007. 3D finite-difference dynamicrupturemodeling along non-planar faults, Geophysics, 72, SM123, doi:10.1190/1.2766756.

Dalguer, L. A., Day S. M., 2007. Staggered-grid split-node method for spontaneous rupture simulation, J. Geophys. Res., 112, B02302, doi:10.1029/2006JB004467.

Das, S., Aki, K., 1977. A Numerical Study of Two-Dimensional Spontaneous Rupture Propagation, Geophysical Journal International 50 (3). Blackwell Publishing Ltd: 643–68. doi:10.1111/j.1365-246X.1977.tb01339.x.

Day, S. M., 1982. Three-Dimensional Finite Difference Simulation of Fault Dynamics: Rectangular Faults with Fixed Rupture Velocity, Bulletin of the Seismological Society of America 72 (3). Seismological Society of America: 705–27.

Day, S. M., Dalguer, L. A., Lapusta, N., Liu, Y., 2005. Comparison of finite difference and boundary integral solutions to three-dimensional spontaneous rupture, Journal of Geophysical Research: Solid Earth, 110(12), 1–23. https://doi.org/10.1029/2005JB003813.

De la Puente, J., Ampuero, J.-P., Käser, M., 2009. Dynamic rupture modeling on unstructured

meshes using a discontinuous galerkin method, J. Geophys. Res., 114, B10302, doi:10.1029/2008JB006271.

Dunham, E. M., Belanger D., Cong L., Kozdon. J. E., 2011. Earthquake Ruptures with Strongly Rate-Weakening Friction and Off-Fault Plasticity, Part 1: Planar Faults, Bulletin of the Seismological Society of America 101 (5): 2296–2307. doi:10.1785/0120100075.

Erickson, B. A., Dunham E. M., Khosravifar, A., 2016. A Finite Difference Method for Off-Fault Plasticity throughout the Earthquake Cycle, Journal of Mechanics and Physics of Solids.

Festa, G., Vilotte J.P., 2006. Influence of the rupture initiation on the intersonic transition: Crack-like versus pulse-like modes, Geophys. Res. Lett., 33, L15320, doi: 10.1029/2006GL026378.

Geubelle, P. H., Rice, J. R., 1995. A Spectral Method for Three-Dimensional Fracture Problems. Journal of the Mechanics and Physics of Solids, 43(11), 1791–1824. https://doi.org/10.1016/0022-5096(95)00043-I

Geubelle, P. H., Breitenfeld, M. S., 1997. Numerical analysis of dynamic debonding under anti-plane shear loading, International Journal of Fracture, 85(3), 265–282. https://doi.org/10.1023/A:1007498300031

Hajarolasvadi, Setare, 2016. A new hybrid numerical scheme for simulating fault ruptures with near fault bulk inhomogeneities, Master's thesis, University of Illinois at Urbana-Champaign, Urbana, http://hdl.handle.net/2142/92863.

Huang, Yihe, Ampuero, J.-P., 2011. Pulse-like Ruptures Induced by Low-Velocity Fault Zones, J Geophys. Res.,116, B12307, doi:10.1029/2011JB008684.

Huang, Y., Ampuero J.-P., Helmberger, D. V., 2014. Earthquake ruptures modulated by waves in damaged fault zones, J. Geophys. Res. Solid Earth, 119, 3133–3154, doi:10.1002/2013JB010724.

Huang, Y., Beroza, G. C. , Ellsworth, W. L. , 2016. Stress drop estimates of potentially induced earthquakes in the Guy-Greenbrier sequence, J.Geophys. Res. Solid Earth, 121, 6597-6607, doi:10.1002/2016JB013067.

Ida, Yoshiaki, 1972. Cohesive Force across the Tip of a Longitudinal-Shear Crack and Griffith's Specific Surface Energy, Journal of Geophysical Research 77 (20): 3796–3805. doi:10.1029/JB077i020p03796.

Kaneko, Y., Lapusta, N., Ampuero, J.-P., 2008. Spectral element modeling of spontaneous earthquake rupture on rate and state faults: Effect of velocity-strengthening friction at shallow depths, J Geophys. Res., 113, B09317, doi:10.1029/2007JB005553.

Käser, M., Dumbser M., 2006. An Arbitrary High-Order Discontinuous Galerkin Method for Elastic Waves on Unstructured Meshes - I. The Two-Dimensional Isotropic Case with External Source Terms, Geophysical Journal International 166 (2). Blackwell Publishing Ltd: 855–77. doi:10.1111/j.1365-246X.2006.03051.x.

Komatitsch, D., Tromp, J., 1999. Introduction to the spectral element method for three-dimensional seismic wave propagation, Geophysical Journal International, 139(3), 806–822. https://doi.org/10.1046/j.1365-246x.1999.00967.x

Kozdon, J.E., Dunham, E.M., Nordström J., 2013. Simulation of Dynamic Earthquake Ruptures in Complex Geometries Using High-Order Finite Difference Methods, Journal of Scientific Computing, 55 (1): 92–124. doi:10.1007/s10915-012-9624-5.

Lapusta, N., Rice, J. R., Ben-Zion, Y., Zheng, G., 2000. Elastodynamic analysis for slow tectonic loading with spontaneous rupture episodes on faults with rate- and state-dependent friction, Journal of Geophysical Research, 105, 23765. https://doi.org/10.1029/2000JB900250

Lysmer, J., Kuhlemeyer, R.L., 1969. Finite dynamic model for infinite media, Journal of the Engineering Mechanics Division, Proc. ASCE, 95 (EM4).

Ma, S., Archuleta, R. J., 2006. Radiated seismic energy based on dynamic rupture models of faulting, J. Geophys. Res., 111, B05315, doi:10.1029/2005JB004055.

Ma, X., Elbanna, A. E., 2015. Effect of off-fault low-velocity elastic inclusions on supershear rupture dynamics, Geophysical Journal International, 203(1), 664-677, doi: 10.1093/gji/ggv302.

Ma, X., Elbanna, A. E., 2017. A Model for Athermal Strain Localization in Dry Sheared Fault Gouge, http://arxiv.org/abs/1701.03087.

Moczo, P., Robertsson, J. O. A., & Eisner, L., 2007. The Finite-Difference Time-Domain Method for Modeling of Seismic Wave Propagation, Advances in Geophysics, 48(6), 421–516. https://doi.org/10.1016/S0065-2687(06)48008-0

Noda, H., Dunham E. M., & Rice, J. R., 2009. Earthquake Ruptures with Thermal Weakening and the Operation of Major Faults at Low Overall Stress Levels, Journal of Geophysical Research 114 (B7): B07302. doi:10.1029/2008JB006143.

Pelties, C., de la Puente, J., Ampuero, J.-P, Brietzke, G., & Käser, M., 2012. Three dimensional dynamic rupture simulation with a high-order Discontinuous Galerkin method on unstructured tetrahedral meshes, J. Geophys. Res., 117, B02309, doi:10.1029/2011JB008857.

Poliakov, Alexei N. B., Dmowska, R., & Rice, J.R., 2002. Dynamic Shear Rupture Interactions with Fault Bends and off-Axis Secondary Faulting, Journal of Geophysical Research 107 (B11): ESE 6-1–ESE 6-18. doi:10.1029/2001JB000572.

Rice, J. R., Sammis, C. G., & Parsons, R., 2005. Off-Fault Secondary Failure Induced by a Dynamic Slip Pulse, Bulletin of the Seismological Society of America 95 (1): 109–34. doi:10.1785/0120030166.

Tago, J., Cruz-Atienza, V. M., Virieux, J., Etienne, V., & Sánchez-Sesma, F. J., 2012. A 3D hp-adaptive discontinuous Galerkin method for modeling earthquake dynamics, J. Geophys. Res., 117, B09312, doi:10.1029/2012JB009313.

Templeton, E. L., & Rice, J. R., 2008. Off-Fault Plasticity and Earthquake Rupture Dynamics: 1. Dry Materials or Neglect of Fluid Pressure Changes, Journal of Geophysical Research: Solid Earth 113 (9): 1–19. doi:10.1029/2007JB005529.

Virieux, J., & Madariaga, R., 1982. Dynamic Faulting Studied by a Finite Difference Method, Bulletin of the Seismological Society of America 72 (2). Seismological Society of America: 345–69.